%% file: main.tex
\documentclass[twoside,11pt]{article}
\usepackage{fullpage}
\usepackage[round]{natbib}

\usepackage[utf8]{inputenc} 
\usepackage[T1]{fontenc}    
\usepackage{hyperref}       

\usepackage{url}            
\usepackage{booktabs}       
\usepackage{amsfonts}       
\usepackage{nicefrac}       
\usepackage{microtype}      

\usepackage{bm}
\usepackage{multirow}

\newcommand{\argmax}{arg\max}

\newcommand{\lightfm}{\textsc{LightFM}}

\newcommand{\pop}{\textsc{POP}}
\newcommand{\warp}{\textsc{WARP}}
\newcommand{\awarp}{\textsc{HMF}}
\newcommand{\nhmf}{\textsc{NHMF}}
\newcommand{\lstm}{\textsc{HA-RNN}}

\newcommand{\cbow}{\textsc{CBOW}}

\newcommand{\bbpr}{b-BPR}

\newcommand{\xing}{\textsf{XING}}
\newcommand{\yelp}{\textsf{Yelp}}

\usepackage{amsmath}
\usepackage{amsfonts}
\usepackage{graphicx}

\usepackage{subfigure}
\usepackage{caption}

\usepackage[
  separate-uncertainty = true,
  multi-part-units = repeat
]{siunitx}

\begin{document}
\title{A Sequential Embedding Approach for Item Recommendation with Heterogeneous Attributes}

\author{
Kuan Liu, Xing Shi, Prem Natarajan\\
 Information Sciences Institute\\
 University of Southern California\\
 \texttt{\{liukuan,xingshi,pnataraj\}@isi.edu }
}
\date{}

\maketitle

\begin{abstract}
Attributes, such as metadata and profile, carry useful information which in principle can help improve accuracy in recommender systems. However, existing approaches have difficulty in fully leveraging attribute information due to practical challenges such as heterogeneity and sparseness. These approaches also fail to combine recurrent neural networks which have recently shown effectiveness in item recommendations in applications such as video and music browsing. To overcome the challenges and to harvest the advantages of sequence models, we present a novel approach, Heterogeneous Attribute Recurrent Neural Networks (HA-RNN), which incorporates heterogeneous attributes and captures sequential dependencies in \textit{both} items and attributes. HA-RNN extends recurrent neural networks with 1) a hierarchical attribute combination input layer and 2) an output attribute embedding layer. We conduct extensive experiments on two large-scale datasets. The new approach show significant improvements over the state-of-the-art models. Our ablation experiments demonstrate the effectiveness of the two components to address heterogeneous attribute challenges including variable lengths and attribute sparseness. We further investigate why sequence modeling works well by conducting exploratory studies and show sequence models are more effective when data scale increases.
\end{abstract}

\input{intro}

\input{attribute}
\input{related}
\input{exp}
\input{conclude}

\bibliographystyle{plainnat}
\bibliography{ref}

\end{document}

%% file: intro.tex

\section{Introduction}
\label{sec:intro}
%

In recent years recommendation with implicit feedback (also known as \textit{item recommendation}) has been increasingly active in research~\cite{hu2008collaborative,rendle2009bpr,johnson2014logistic,he2016fast}, and is applied in various applications like e-commerce~\cite{linden2003amazon}, social networks~\cite{chen2009make}, location~\cite{liu2014exploiting}, etc. After observing user activities such as clicks, page views, and purchases---often called implicit feedback---the goal is to recommend to each user a ranked list of items that he might prefer. Item recommendation remains a very challenging task. Most users have very limited interactions with extremely small portions of items and often no negative feedback~\cite{hu2008collaborative}. User-item interactions are implicit and hard to interpret~\cite{hu2008collaborative,liu2015boosting}. Different collaborative filtering methods are heavily exploited to alleviate these challenges~\cite{hu2008collaborative,rendle2009bpr,rendle2014improving,usunier2009ranking,weston2010large,shi2012tfmap,lirelaxed}.

Content-based approaches are important alternative methods to collaborative filtering. Increasingly prevalent attributes (such as metadata or profile) of users and items, e.g., LinkedIn user profiles and Yelp business metadata, offer new opportunities and trigger active research. To make use of the attributes, low-rank factor models such as matrix co-factorization~\cite{fang2011matrix,saveski2014item}, tensor factorization~\cite{karatzoglou2010multiverse,bhargava2015and}, and hybrid matrix factorization~\cite{shmueli2012care,kula_metadata_2015} are developed to jointly learn latent factors of attributes as well as users and items. While these matrix factorization extensions provide a simple mechanism to incorporate attributes, in practice they are not adequate to fully leverage attribute information due to challenges such as heterogeneity and sparseness. Moreover, these approaches are based on bi-linear models and thus are too restrictive in their model flexibility to capture complex dependencies between different model components.

\begin{figure}[!h]
\centering
\caption{An illustrating example where a job seeker interacts with a sequence of job posts. Rich heterogeneous attributes are associated with both the job seeker and job posts. The system is asked to recommend new posts to this user.}
\label{f:example}
\includegraphics[width=1.0\columnwidth]{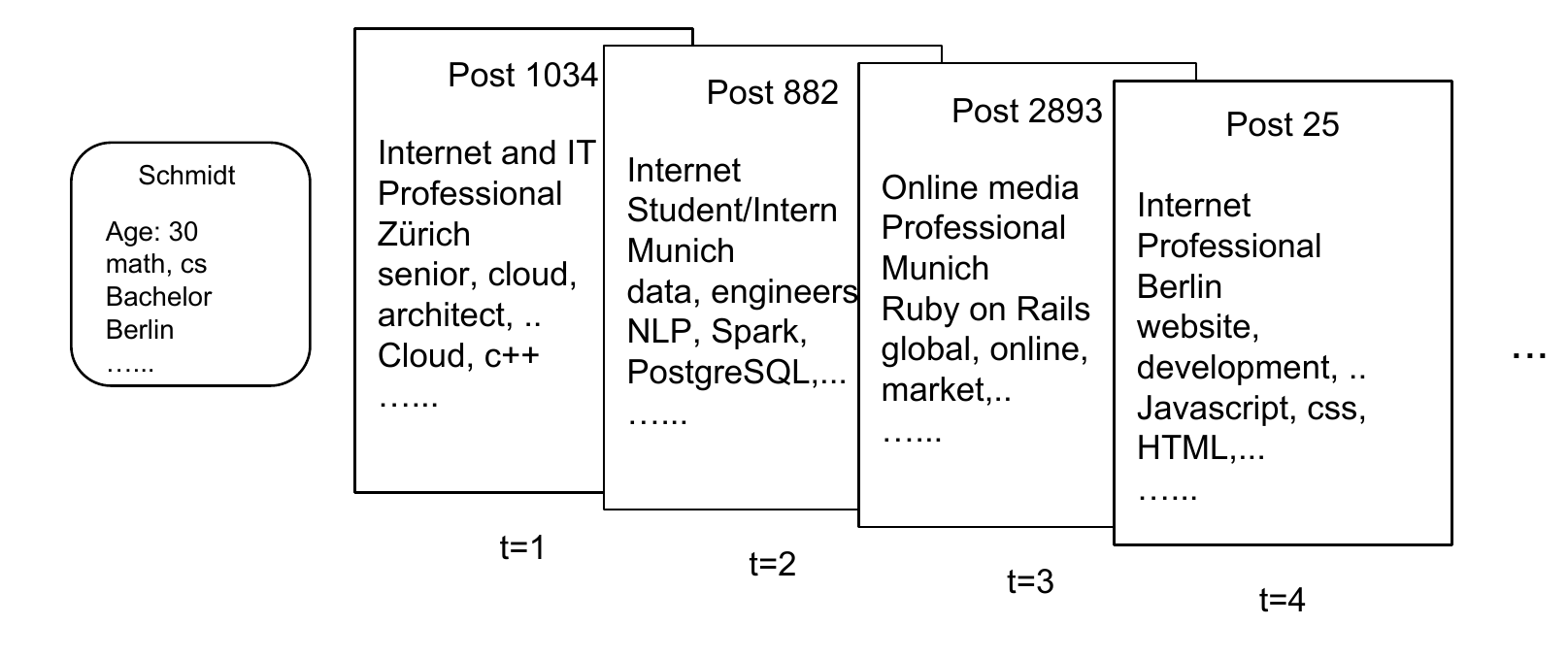}
\end{figure}



Sequential dependency is a prominent example of such dependencies. Item sequential dependencies have proved useful in recommender systems~\cite{rendle2010factorizing,cheng2013you,kapoor2015just,du2015time}. Recently, flexible sequence models based on RNNs and Word2Vec are used to model item transitions explicitly and show better prediction performances than factor models in E-commerce~\cite{grbovic2015commerce}, music~\cite{vasile2016meta}, and videos~\cite{hidasi2015session}. However, these sequence models either work with no attributes or one or two domain-specific features. Dependencies between attributes and items and within attributes are largely ignored. This also prevents better understanding of user behaviors and further improvement to system performance.

Attributes are often heterogeneous and come in different domains and data types. For example (as in Fig.\ref{f:example}), attributes may include a user's age, location, education, and a job's industry, title description, employment, online tags; their data types include real numbers, categorical tokens, text tokens, etc. To develop efficient approaches to incorporating attributes, we investigate three major challenges in real scenarios---variable lengths, sparseness, and sequential dependency. {Variable lengths}: Due to the natures of different attributes, users or items may have different lengths of attributes. A user may have academic degrees in more than one discipline; a job post may vary in the number of tags; missing values are very common. It is inefficient to simply combine the observed attributes. {Sparseness}: The possible attributes of items include tags from the Internet, text tokens, demographic information, etc. The entire attribute vocabulary can be very large, but each attribute appears only a limited number of times. To add difficulty, a large part of the attributes may be irrelevant to our task of interest. These factors pose further challenges to model regularization. {Sequential dependency}: In cases where a user interacts with a sequence of items (or services), attributes may help encode user sequential behaviors. As evidence, attributes of nearby items in a sequence may overlap or share common characteristics. For example as in Fig.\ref{f:example}, job post attributes collectively suggest the user's current interest in a junior-level position in an IT-related area. It is desirable to make use of such attribute dependencies.

In this work we propose a novel model, Heterogeneous-Attribute Recurrent Neural Networks ({HA-RNN}), to address the above challenges. HA-RNN combines sequence modeling and attribute embedding in item recommendation. Different from conventional RNNs, HA-RNN develops novel embedding techniques and integrates them into sequence modeling. In particular,
\begin{itemize}
\item It develops a hierarchical attribute combination mechanism to deal with variable lengths of attributes. 
\item To battle against attribute sparseness, the model uses attributes in the output layer and shares the parameters with the input layer to offer additional model regularization.
\item HA-RNN takes the union of identity and attributes as a sequence element and is able to capture the global sequential dependencies between items as well as between attributes.
\end{itemize}
The model is trained end-to-end and learns task-driven attribute representations

Experiments are conducted on two large-scale datasets for implicit recommendation. Our methods give significant improvement compared to state-of-the-art baseline models. Results show that attribute embedding and sequence modeling are both essential to improving performance. Particularly, the novel output attribute layer design helps boost recommendation accuracy. We qualitatively demonstrate our model's ability to discover attribute semantic structures. Finally, further analysis is conducted to understand how item sequences influence recommendation.

The contributions of this work are summarized as follows: 
\begin{itemize}
\item We study the problem of item recommendation with heterogeneous attributes. Challenges to incorporate these attributes include variable lengths, sparseness, and sequential dependencies. 
\item We propose an approach that combines attribute embedding and recurrent neural networks to incorporate attribute in item recommendation. We develop novel techniques to address attribute heterogeneity challenges. 
\item We perform empirical studies on two real-world datasets and show the effectiveness of our approach. Our model significantly outperforms the state-of-the-art models. Detailed analysis indicates the critical roles of our different model components. 
\end{itemize}

%% file: attribute.tex
\section{Approach}
\label{sec:attribute}

\begin{figure*}[!h]
\centering
\caption{The overview of the proposed HA-RNN model. The base recommendation model is recurrent neural networks where item sequence (with the user) is fed as input and the model is trained to predict the next item. The representations for the input  ($\bm{Q}_U$ and $\bm{Q}_I$) and for computing item scores in the output layer ($\bm{Q}_I$) are combination of identity and attributes. To compute $\bm{Q}_I$, embedding of the multi-hot attributes (e.g., $\bm{\phi}_{M_1}$) are first averaged before combined with categorical ones (e.g., $\bm{\phi}_{C_1}, \bm{\phi}_{C_2}$; identity is regarded as categorical attribute in the figure) and numerical ones (e.g., $\bm{\phi}_{N_1}$). The same item representation is shared in both input and output layers for enhanced signals and model regularization. Computation of $\bm{Q}_U$ is omitted in the figure.}\label{pic:overview}
\label{f:overview}
\includegraphics[width=1.0\columnwidth]{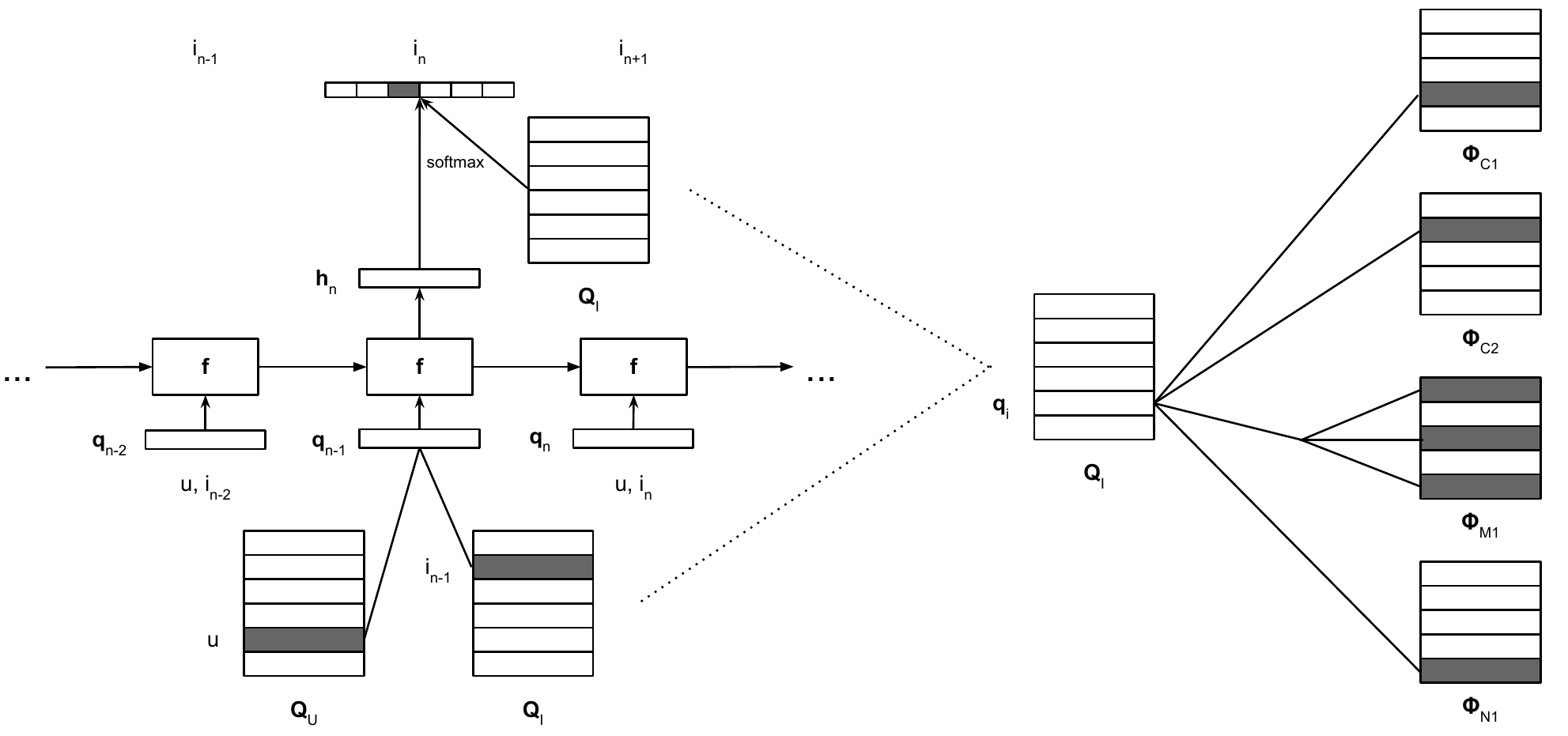}
\end{figure*}


In this section we introduce our proposed model HA-RNN. After defining the problem formulation and notation, we first present an existing approach based on RNNs in item recommendation. Then we describe our proposed model that combines sequence modeling and attribute embedding and deals with heterogeneity challenges.  

\subsection{Problem Formulation and Notation}

We are given a user set $U$, an item set $I$, and their interactions $S=\{(u,i,t) | u\in U, i\in I\}$ where $t$ records the time that interaction $(u,i)$ takes place. Let $A^U,$ $A^I$ denote the attribute set associated with users and items. Parameters associated with users, items, user attributes, and item attributes are denoted by $\bm{e}^{U}\in\mathbb{R}^{|U|\times d}$, $\bm{e}^{I}\in\mathbb{R}^{|I|\times d}$, $\bm{\phi}^{U}\in\mathbb{R}^{|A^U|\times d}$, and $\bm{\phi}^{I}\in\mathbb{R}^{|A^I|\times d}$, respectively--- given the model dimensionality $d$. Subscripting means taking one corresponding slice of the matrix. For example, $\bm{e^I}_i\in\mathbb{R}^d$ denotes a vector representation of item $i$; $\bm{\phi^U}_a\in\mathbb{R}^d$ ($a\in A^U$) denotes embedding of one user attribute $a$. Superscript $U$ or $I$ is omitted when there is no ambiguity. In the context of a sequence, integer subscripts denote sequence positions.

A recommender system needs to return a scoring function g such that $s(i) = g(i |u, U,I, S, A^U, A^I)$ captures user-item preferences where an item preferred by a user to another receives a higher score. The recommendation then takes the highest scored item (as shown in Eq.~\ref{eq:pred}).



\begin{equation}
\label{eq:pred}
\hat{y} = \argmax_{i\in I} s(i)
\end{equation}

In model training the scores are compared to the ground truth observations. Different loss functions are developed~\cite{hu2008collaborative,rendle2009bpr,weston2010large}. In this work we use cross-entropy classification loss as in~\cite{grbovic2015commerce,hidasi2015session}.

\subsection{Identity Embedding via Sequential Recommendation}
\label{sec:itemseq}
We begin with an existing sequence recommendation approach \cite{hidasi2015session}. First, items seen by a user $u$ are sorted chronologically as a sequence $(u:i_1,i_2,..)$. Then, as in a sequence generation problem, an RNNs-based model learns identity embedding of users and items by fitting the model to predict the next appearing item given all the observed ones. The model formulation can be written as the following,
\begin{align}
    &\bm{h}_n = f(\bm{q}_{n-1}, \bm{h}_{n-1}) 
   \quad  \forall n = 1,..,t  \label{eq:lstm_forward} \\
   &s_n({i_{n}}) = \bm{h}_n^T\bm{w}_{i_{n}}  \label{eq:score}
\end{align}
where $f$ is an \text{RNN-cell} (e.g., LSTM, GRU); $s({i_n})$ is computed from the inner product between latent state $\bm{h}_n$ and item matrix $\bm{w}\in\mathbb{R}^{d\times |I|}$.
\footnote{A special ``START'' symbol is used as the very first item, i.e., $\bm{q}_{i_0}$, and $\bm{h}_0= \bm{0}$.}


In this work we focus on $\bm{q}$ and $\bm{w}$. As the input layer and the output layer of networks, they connect sequence models with data observations. In \cite{hidasi2015session}, the input vector $\bm{q}$ simply takes item embedding,
\begin{equation}
\bm{q}_n  = \bm{e}_{i_n},
\end{equation}
and $\bm{w}$ is a separate set of item parameters from $\bm{e}^I$. While the model is able to capture sequential dependencies between items, it does not use attributes at all as neither $\bm{q}$ nor $\bm{w}$ involve attributes.

%


\subsection{Joint Attribute Embedding}
\label{sec:combine}
We first extend the model to incorporate attributes in the input layer---both item identity and attributes contribute to the input representation. Particularly, 

\begin{equation}
\label{eq:combine}
\bm{q}_i = \bm{e}_i + \sum_{j\in attr(i)} \bm{\phi}_j
\end{equation}
where $attr(i)$ returns the set of attributes of item $i$. When user attributes are also available, we similarly define $\bm{q}_u$ and thus have $\bm{q}_n=\bm{q}_u+\bm{q}_{i_n}$.



This model, however, is inefficient in practice. It suffers from the mentioned challenge---variable attribute lengths. For example, when an item has more attributes than others, its second summand in Eq.~\ref{eq:combine} tends to have larger magnitude, and the estimation becomes harder. As another example, if an item has more attributes for one type but less for another, the mismatch is harmful when the model tries to compare representations of two items.



\subsection{A Hierarchical Attribute Combination}
\label{sec:mulhot}

\begin{table}
\centering
\caption{An attribute division example. How attributes in Figure \ref{f:example} are divided into three kinds.}
\label{t:division}
\begin{tabular}{cc} \hline
Divisions & Examples \\ \hline
Categorical & industry(Online media), location(Berlin)... \\ \hline
Multi-hot & title(data, market), tags(NLP, Spark)...\\ \hline
Numerical & age(30)\\ \hline
\end{tabular}
\end{table}

To deal with the challenge, we propose a division of heterogeneous attributes into three distinct kinds: \textit{categorical} ($\mathcal{C}$), \textit{multi-hot} ($\mathcal{M}$), and \textit{numerical} ($\mathcal{N}$). In particular, 
\begin{itemize}
\item \textit{Categorical} attributes are those with exclusive labels, which means an object can only have one value for that type. 
\item \textit{Multi-hot} attributes are often descriptive, and an object is accompanied by one or more attribute values of a type. 
\item \textit{Numerical} attributes can have continuous values in real numbers with their specific domain interpretations. 
\end{itemize} 

In the example in Figure \ref{f:example}, \textit{categorical} attributes include the job post identity, industry, employment requirement, and location. Title text tokens, tags here or in other examples like movies tags are \textit{multi-hot} attributes.  User ``age'' here belongs to the \textit{numerical} group.

We point out that \textit{multi-hot} attributes---which can have more than one value for each attribute---and missing values are the causes of variable lengths of attributes.\footnote{Missing values can be replaced by ``unk'' as special attribute tokens.} 


With the division and a bit of abuse of notation $\mathcal{C},\mathcal{M},\mathcal{N}$, we design a hierarchical way to combine the attributes and have the representation as follows:
\begin{equation}
\label{eq:item_input}
\bm{q}_i = \bm{e}_i + \sum_{j=1}^{n_\mathcal{C}}\bm{\phi}_{\mathcal{C}_j(i)} + \sum_{j=1}^{n_\mathcal{M}}\frac{1}{|\mathcal{M}_j(i)|}\sum_{k\in \mathcal{M}_j(i)}\bm{\phi}_k + \sum_{j=1}^{n_\mathcal{N}}\bm{\phi}_{\mathcal{N}_j(i)}
\end{equation}
where $\mathcal{C}_j(i)$ (or $\mathcal{M}_j(i),\mathcal{N}_j(i)$) returns item $i$'s $j$th categorical (or multi-hot, numerical) attribute(s).\footnote{\textit{Numerical} attribute values are first clustered and then replaced by their cluster center indices. They are then treated as \textit{categorical} values.} $n_{\mathcal{C}}$ (or $n_{\mathcal{M}},n_{\mathcal{N}}$) denotes the total number of attribute types belonging to $\mathcal{C}$ (or $\mathcal{M},\mathcal{N}$). $\bm{q}_u$ is computed similarly.

Compared to (\ref{eq:combine}), multi-hot attribute embedding is no longer summed together. Rather, the mean of embedding within each type of multi-hot attribute is computed before embeddings across types are combined. In this way, values of one multi-hot attribute are regarded as a ``single attribute'', and we have a better control of the scales of input vector $\bm{q}$.

\subsection{Shared Attribute Embedding in Output Layer}

With (\ref{eq:lstm_forward})(\ref{eq:score})(\ref{eq:item_input}), RNNs embed attributes as part of model input. However, as most conventional RNNs, attributes are not involved in the output prediction stage. To address the attribute sparseness challenge, we first explore incorporating attributes in the output layer in addition to the input layer. Intuitively, use of attributes in these two network components enhances attribute signals in model training.

We extend the output layer to involve attribute embedding in computing item scores. Particularly, we discard $\bm{w}$ and compute the score for item $i$ by

\begin{align}
\label{eq:output_layer}
\begin{split}
s(i) = \bm{h}^T\bm{e}'_i &+ \sum_{j=1}^{n_C}\bm{h}^T\bm{\phi'}_{C_j(i)} + \sum_{j=1}^{n_M}\oplus_{k\in M_j(i)}\bm{h}^T\bm{\phi'}_k \\
&+ \sum_{j=1}^{n_N}\bm{h}^T\bm{\phi'}_{N_j(i)} \qquad \forall i\in I
\end{split}
\end{align}
where $\oplus\in\{mean,max\}$. First, attribute preference scores are computed by dot product between model latent vector $\bm{h}$ and attribute vectors. Then the attribute scores within each \textit{multi-hot} type take \textit{mean} or \textit{max}. Intuitively, \textit{mean-pooling} suggests that each attribute value contributes equally, while \textit{max-pooling} suggests that one particularly favorable attribute value dominates. Finally, scores across different attribute types are summed (or averaged) to produce item scores.



Note that item identity and attributes now appear both in input and output layers of the model. A natural question is whether or not we should use separate embedding parameters for the two components. Different from common practice such as in Word2Vec, we decide the input and output layers share the same embedding parameters, i.e.,
\begin{equation}
\label{eq:share}
\bm{e}' = \bm{e}, \qquad \bm{\phi}' = \bm{\phi}.
\end{equation}

This reduces the number of total parameters and, more importantly, adds additional constraints to embedding parameters. We expect this model regularization to help the model achieve better generalization. 

\textbf{HA-RNN}. With (\ref{eq:lstm_forward})(\ref{eq:score})(\ref{eq:item_input})(\ref{eq:output_layer})(\ref{eq:share}), we have our complete model HA-RNN. As in Figure \ref{f:overview}, attributes and identities are coupled in the sequence model through both input and output layers. Given an item sequence, the model first looks up and combines the attribute embeddings and then feeds the representation into the networks. Attribute parameters are updated together with network parameters via back-propagation. Compared to the sequential recommendation model in Sec.~\ref{sec:itemseq}, HA-RNN treats the union of identity and attributes as a sequence element and tries to capture sequential dependencies in \textit{both} identities and attributes. Compared to simple attribute incorporation in Sec.~\ref{sec:combine}, HA-RNN carefully designs input and output layers to deal with heterogeneity challenges.

%% file: related.tex
\section{Related Work}
\label{sec:related}
\textbf{Attribute incorporation.}
Low rank factorization models are extended to incorporate attributes in recommendation. Matrix co-factorization (MCF)~\cite{fang2011matrix,saveski2014item} minimizes a sum of several matrices' losses to capture multiple data relation. Hybrid matrix factorization (HMF)~\cite{shmueli2012care,kula_metadata_2015} uses linear combination of attribute embedding to represent use or item and then factorize only one interaction matrix. Tensor factorization~\cite{karatzoglou2010multiverse,bhargava2015and}~models attributes as part of tensors. These approaches are based on bi-linear models and are often limited in the model expressiveness and flexibility.

Recently, more flexible models (e.g., topic models and \textit{pre-trained} neural networks) have been used to model attributes of text~\cite{bansal2016ask,kim2016convolutional}, vision~\cite{he2015vbpr}, and music~\cite{van2013deep}. The learned embedding from these flexible models are then fixed and used in a downstream matrix factorization recommender system. As a comparison, our approach targets  heterogeneous types of attributes; meanwhile, it learns attribute embedding in a task-driven fashion to avoid domain adaption issues.

Graph-based algorithms like graph random walk are used to learn attributes~\cite{christoffel2015blockbusters,levin2016guided}. They construct graph nodes as identities and attributes, and graph edges as events and attribute relations.  Attributes and identities are then embedded in the same space.

\textbf{Sequence modeling.}
Sequence models have proved useful in recommender systems. Low rank factorization has been effectively combined with Markov chain~\cite{rendle2010factorizing,cheng2013you}, Hidden Semi-Markov model~\cite{kapoor2015just}, Hawkes processes~\cite{du2015time}.

Recently, Word2Vec training techniques are tailored to recommendation models. It shows scalability and efficiency in E-commerce product recommendation~\cite{grbovic2015commerce}.~\cite{vasile2016meta} extends~\cite{grbovic2015commerce} to incorporate metadata in a multitask learning fashion and shows improved performance on a music dataset. 

\cite{hidasi2015session} first applies recurrent neural networks (RNNs) to session-based recommendation. It devises GRU-based RNNs and demonstrates good performance with one hot encoding item embedding. \cite{liu2016temporal} explores LSTMs in general recommendation domains on an online job recommendation task. \cite{hidasi2016parallel} extends \cite{hidasi2015session} by building a parallel structure to take in visual extracted features in the input layer. \cite{liu2016context} extends recurrent neural networks in context-aware recommendation by introducing context-dependent network components.

\textbf{Comparisons}
In designing our approach to incorporate attributes, we have the following considerations: 1) task-driven: attribute embedding is learned end-to-end for recommendation tasks; 2) heterogeneous: the model can handle more than a few attribute types;  3) flexible in model capability: we want the model to go beyond matrix factorization; 4) sequential: it has the ability to model sequential dependencies. We compare our approach to other related works in Table \ref{t:related} and highlight our approach combining all these properties. While we do not argue these properties are in general the most desirable, we find in our empirical study they are helpful to guide to achieve good performance.

\begin{table}
\caption{Comparisons of different attribute incorporation methods.}
\centering
\label{t:related}
\begin{tabular}{|c|c|c|c|c|} \hline
Methods & Task-driven & Het. & Flexible & Sequential \\ \hline
MCF, Multi-task& & $\surd$ &  &\\ \hline
HMF, Tensor-based  & $\surd$ & & & \\ \hline
Pre-trained & & & $\surd$ & $\surd$ \\ \hline
Graph-based & $\surd$ & $\surd$ &  &  \\ \hline
Ours & $\surd$ & $\surd$ & $\surd$ & $\surd$ \\ \hline
\end{tabular}
\end{table}

%% file: exp.tex
\section{Experiments}
\label{sec:exp}

This section discusses how we conduct extensive experiments to validate the effectiveness of our approach. First, we introduce the experimental setup. Then we present the recommendation accuracy results comparing different variants of HA-RNN and compare these with other state-of-the-art models. This is followed by a qualitative analysis of learned attribute embedding. Finally, we review our studies on how the sequence component helps recommendation. The results demonstrate how the embedding techniques and sequence modeling help achieve recommendation accuracy improvement.

\subsection{Experimental Setup}
\subsubsection{Datasets}
We validate our approach on two large-scale datasets: 1) job recommendation from XING\footnote{www.xing.com}; 2) business reviews at Yelp. Both datasets contain rich attributes associated with users and items, in addition to user-item interactions. We describe the datasets in detail below.

\textbf{XING} is a dataset adapted from the RecSys Challenge 2016 \cite{abel2016recsys}. It contains about 14 weeks of interaction data between 1,500,000 users and 327,002 job posts on XING. The interactions are user online activities including \textit{click}, \textit{bookmark}, and \textit{reply}. Train/test splitting follows the official challenge setup where the interactions in the first 12 weeks are used as training data, and the interactions for a set of 150,000 \textit{target users} in weeks 13 and 14 are used as test data. Our task is to recommend a list of job posts to the target users; these are the posts that  users are  likely to interact with (either \textit{click}, \textit{bookmark}, or \textit{reply}) in the last two weeks. 

Rich attributes are associated with users and job posts. Attributes are anonymized tokens in different fields such as career levels, education disciplines, industries, locations, work experiences, job roles, job titles, etc. For example, an attribute token may represent a HR discipline, or consulting; it may also represent a word in a job title. The detailed attribute information is listed in Table~\ref{t:data2}. We remove attributes that appear less than two times. The size of the remaining attribute vocabulary is 57,813.


\textbf{Yelp} is a dataset that comes from the Yelp Challenge.\footnote{https://www.yelp.com/dataset\_challenge. Downloaded in Feb 17.} There are 1,029,433 users and 144,073 businesses. Interactions here are defined as \textit{reviews}. We work on recommendation related to which business a user might want to review. Following the \textit{online protocol} in \cite{he2016fast}, we sort all interactions in chronological order, taking the last 10\% for testing and the rest for training. There are 1,917,218 reviews in the training split and 213,460 in the testing split. 

Business items have attributes including \textit{city}, \textit{state}, \textit{categories} (e.g., ``meditation centers,'' ``thai foods''),  \textit{hours}, and \textit{features} (e.g., ``valet parking,'' ``good for kids''). Similarly, we remove rare attributes with less than two appearances. The attribute vocabulary size is 37,573. Dataset statistics are listed in Table~\ref{t:data1}.

\begin{table}
\centering
\caption{\xing~attribute descriptions (U: user; I: item).}
\begin{tabular}{|c|c|}
\hline
\textbf{Feature types}  & \textbf{Features} \\ \hline
\multirow{3}{*}{Categorical (U)} & career\_level, discipline\_id, industry\_id,\\
  &  user\_id, country, region, exp\_years,\\
& exp\_in\_entries\_class, exp\_in\_current\_job\\ \hline
Multi-hot (U) & job\_roles, field\_of\_studies \\ \hline
\multirow{2}{*}{Categorical (I)} & item\_id, career\_level, discipline\_id, \\
& country, region, employment \\ \hline
Numerical (I) & latitude, longitude, created\_at \\ \hline
Multi-hot (I) & title, tags \\ \hline 
\end{tabular}
\label{t:data2}
\end{table}

\begin{table}
\centering
\caption{Statistics of data sets.}
\label{t:data1}
\begin{tabular}{|c|c|c|c|c|}
\hline
\textbf{Data}   & $\left|U\right|$ & $\left|I\right|$ & $\left|S_{\text{train}}\right|$ & $\left|S_{\text{test}}\right|$ \\
\hline
\xing       & 1,500,000     & 327,002  & 2,338,766  & 484,237 \\ \hline
\yelp       & 1,029,433     & 144,073  &  1,917,218 & 213,460  \\ \hline
\end{tabular}

\end{table}

\subsubsection{Evaluation}
We assess the quality of recommendation results by comparing the models' recommendations to ground truth interactions, and report \textit{precision}, \textit{recall}, \textit{MAP}, and \textit{normalized discounted cumulative gain  (NDCG)}. For precision and  recall~metrics, we report results at positions 2, 10, and 30. We report MAP and NDCG at location 30. Note that we report scores \textit{after removing historical items from each user's recommendation list} on the~\yelp~dataset because in these scenarios (Yelp reviews), users seldom re-interact with items. This improves performance for all models, but does not change relative comparison results.

\begin{table*}[t]
\caption{Recommendation accuracy and training time comparisons between MIX and HET on different models.}
\label{t:mulhot}
\centering
\begin{tabular}{cccccccccc} \hline
 Attr. & Time & P@2 & P@10 & P@30 & R@2 & R@10  & R@30& MAP & NDCG \\ \hline
 \multicolumn{10}{c}{\xing} \\ \hline
 MIX& $>$144h & 0.0452 & 0.0221 & 0.0122 & 0.0396 & 0.0891 & 0.1394 & 0.0486 & 0.0806\\
 HET  & 70h & \textbf{0.0522} & \textbf{0.0241} & \textbf{0.0128} & \textbf{0.0451} & \textbf{0.0962} & \textbf{0.1453} & \textbf{0.0540} & \textbf{0.0873}\\ \hline
 \multicolumn{10}{c}{\yelp} \\ \hline
MIX& $>$144h& 0.0152 & 0.0119 & 0.0090 & 0.0052 & 0.0204 & 0.0469 & 0.0093 & 0.0265\\
HET & 82h & \textbf{0.0205} & \textbf{0.0154} & \textbf{0.0117} & \textbf{0.0080} & \textbf{0.0281} & \textbf{0.0638} & \textbf{0.0131} & \textbf{0.0357} \\ \hline
\end{tabular}
\end{table*}


\begin{table*}[t]
\caption{Recommendation accuracy comparisons of different attribute embedding integrations.}
\label{t:output}
\centering
\begin{tabular}{ccccccccc} \hline
Methods & P@2 & P@10 & P@30 & R@2 & R@10  & R@30& MAP & NDCG \\ \hline
\multicolumn{9}{c}{\xing} \\ \hline
no attributes 				& 0.0406 & 0.0169 & 0.0091 & 0.0335 & 0.0597 & 0.0908 & 0.0365 & 0.0587\\
only input 			& 0.0499 & 0.0233 & 0.0126 & 0.0431 & 0.0924 & 0.1414 & 0.0516 & 0.0842\\
only output 			& 0.0515 & 0.0236 & 0.0127 & 0.0445 & 0.0936 & 0.1427 & 0.0530 & 0.0858\\
input+output 			& \textbf{0.0522} & \textbf{0.0241} & \textbf{0.0128} & \textbf{0.0451} & \textbf{0.0962} & \textbf{0.1453} & \textbf{0.0540} & \textbf{0.0873}\\  \hline
%
\multicolumn{9}{c}{\yelp} \\ \hline
no attributes 				& 0.0183 & 0.0138 & 0.0104 & 0.0069 & 0.0248 & 0.0551 & 0.0113 & 0.0313\\
only input 			& 0.0175 & 0.0138 & 0.0106 & 0.0063 & 0.0248 & 0.0560 & 0.0111 & 0.0314\\
only output 			& 0.0180 & 0.0138 & 0.0104 & 0.0067 & 0.0250 & 0.0557 & 0.0112 & 0.0313\\
input+output 			& \textbf{0.0205} & \textbf{0.0154}& \textbf{0.0117} & \textbf{0.0080} & \textbf{0.0281} & \textbf{0.0638} & \textbf{0.0131} & \textbf{0.0357} \\  \hline
\end{tabular}
\end{table*}

\subsubsection{Models}
We compare our models with models that show variant abilities in incorporating attributes and utilizing sequence in item recommendation. Baseline non-sequence models~\pop~and~\warp~\cite{weston2010large} and sequence models~\cbow~\cite{grbovic2015commerce}~and~RNN~\cite{hidasi2015session}~do not incorporate attributes. Non-sequence model~\awarp~\cite{kula_metadata_2015}~and sequence approach A-RNN incorporate attributes. We also experimented with different variants of our proposed HA-RNN to investigate the roles of different model components. We detail the models and hyper-parameter tuning below. Hyper-parameters tuning and early stopping are done on a development dataset split from training for all models. Selected hyper-parameters do not land on tuning boundaries.

\begin{itemize}
\item \pop. A naive baseline model that recommends items in terms of their popularity. 
\item \warp. One state-of-the-art algorithm for item recommendation \cite{weston2010large,hong2013co}. We use~\lightfm~implementation~\cite{kula_metadata_2015}. Major hyper-parameter model dimension $d$ is tuned in \{16, 32, 48, 64\}. 
\item \cbow. A sequence recommendation approach ~\cite{grbovic2015commerce} based on Word2Vec techniques. We used our own implementation. We use window size 5, dropout rate 0.5. $m$ in \cbow~is tuned in \{1,2,3,4\}. We also experimented with its alternative approach based on Skip-gram. \cbow~works better. We omit our results for brevity.
\item RNN. A sequence approach~\cite{hidasi2015session} based on RNNs. We used our own implementation. Model dimension $d$ is tuned in \{64, 128, 256\}. Dropout rates are tuned between 0.3 and 0.8.
\item \awarp. A factorization model that represents users and items as linear combinations of their attribute embedding~\cite{shmueli2012care}. We use~\lightfm~implementation~\cite{kula_metadata_2015}. Model dimension $d$ is tuned in \{10, 16, 32, 48, 64\}.
\item \nhmf. Our own implemented HMF \cite{liu2017wmrb}\cite{liu2017batch}. It uses cross-entropy as a training algorithm instead of the rank loss used in \awarp. AdaGrad is used to optimize cross-entropy loss. Dropout rate 0.5. Model dimension $d$ is tuned in \{16, 32, 48\}. 
\end{itemize}

\subsection{Recommendation Accuracy}

\subsubsection{Effectiveness of multi-hot treatment}
To begin, we investigate the effectiveness of multi-hot treatment in our attribute embedding. We call our combination (\ref{eq:item_input}) discussed in Section \ref{sec:mulhot} \textbf{HET}. We compare this to (\ref{eq:combine}) (we call \textbf{MIX}) which simply takes the average embedding of all attribute values. Recommendation accuracies and training time are reported in Table \ref{t:mulhot}.

In terms of recommendation accuracy, our multi-hot treatment HET brings significant improvement to HA-RNN. In both datasets, HET significantly outperforms MIX. Particularly in \yelp, the improvement is dramatic (41\% relative gains on MAP and 34\% on NDCG). This may be due to the fact that variable lengths of attributes are really harmful to flexible models such as RNNs. Our HET deals with this heterogeneity and provides within-category normalization. This turns out to be very critical in model regularization. 

In terms of training time, we see that it takes much longer to train HA-RNN with MIX than HET. This is expected as the MIX model has to additionally adjust the embedding scales while HET does not. We stopped the training process of HA-RNN-MIX after 6 days.

\subsubsection{Effectiveness of output layer}

To evaluate the effectiveness of the output attribute embedding layer, we explore four variants of \lstm: 1) `no attributes': No attributes, but identity is used---this degenerates to a recommendation model based on standard RNNs as in \cite{hidasi2015session}; 2) `only input': attributes used at input layer; 3) `only output': attributes used at output layer; 4) `input+output': attributes used at both input and output layers. 
Experiment results are reported in Table~\ref{t:output}, from which we make two observations.

First, \lstm~performs poorly on both datasets when no attributes are used (`no attributes'). On \xing, the result is even worse than that of non-sequence model NHMF (see \nhmf~HET variant in Table \ref{t:mulhot}). On the contrary, with attribute embedding (`input+output'), the \lstm~scores are significantly boosted. These results suggest that sequence modeling alone is not good enough, and attributes do help to improve accuracy.

Second, the output layer design for attribute embedding turns out to be very important. On \xing, only incorporating attributes at the output layer (`only ouput') gives better results than only at the input layer (`only input'). Moreover, it works best to learn attributes at both layers (`input+output'). On \yelp, embedding attributes either at input layer or output layer alone seems hardly helpful; however, by embedding attributes at both layers we manage to significantly improve the score. This verifies our assumption that the output layer embedding brings additional regularization.


To validate the effectiveness of sharing attribute parameters in both input and output embedding layers, we run HA-RNN with separate sets of parameters $\bm{\phi}, \bm{e}$ and $\bm{\phi}', \bm{e}'$. We report validation perplexities in Table \ref{t:shared}. `only output' and `in+out (shared)' are models we have discussed in Table \ref{t:output}. `in+out (sep)' denotes the new model configuration. 

In Table \ref{t:shared}, `in+out (sep)' clearly underperforms `in+out (shared)' in both datasets; on~\xing~dataset, it is even worse than `only output' model. Additional attribute parameters in the output layer lead to worse generalization. This suggests the necessity of sharing parameters across input and output layers.

\begin{table}
\centering
\caption{Validation perplexity comparisons of different output layer attribute configurations.}
\label{t:shared}
\begin{tabular}{|c|c|c|c|} \hline
Datasets & only output & in+out (sep) & in+out (shared) \\ \hline
\xing & 1201 & 1495 & \textbf{1115} \\ \hline
\yelp & 4721 & 4131 & \textbf{3658} \\ \hline
\end{tabular}
\end{table}


\subsubsection{Performance comparisons}

\begin{table*}[t]
\centering
\caption{Recommendation accuracy comparisons with other state-of-the-art models. Best and second best single model scores are in bold and italic, respectively.}
\label{t:stoa}
\begin{tabular}{ccccccccc} \hline
 Methods & P@2 & P@10 & P@30 & R@2 & R@10  & R@30& MAP & NDCG \\ \hline
 \multicolumn{9}{c}{\xing}
  \\ \hline
 \pop   		  & 0.0077 & 0.0034 & 0.0021 & 0.0079 & 0.0154 & 0.0274 & 0.0062 & 0.0127\\ 
 \bbpr                & 0.0159 & 0.0111 & 0.0083 & 0.0133 & 0.0423 & 0.0920 & 0.0197 & 0.0420 \\
 \warp 		  & 0.0368 & 0.0179 & 0.0092 & 0.0309 & 0.0608 & 0.0832 & 0.0346 & 0.0557\\ 
 CBOW 		  & 0.0367 & 0.0164 & 0.0088 & 0.0298 & 0.0571 & 0.0870 & 0.0329 & 0.0543\\
 RNN 		  & \textit{0.0406} & 0.0169 & 0.0091 & 0.0335 & 0.0597 & 0.0908 & 0.0365 & 0.0587\\
 \awarp		  & 0.0362 & \textit{0.0194} & \textit{0.0109} & \textit{0.0325} & \textit{0.0749} & \textit{0.1163} & \textit{0.0398} & \textit{0.0674}\\ 
 HA-RNN 	  & \textbf{0.0522} & \textbf{0.0241} & \textbf{0.0128} & \textbf{0.0451} & \textbf{0.0962} & \textbf{0.1453} & \textbf{0.0540} & \textbf{0.0873}\\ 
 HA-RNN* & 0.0537 & 0.0252 & 0.0134 & 0.0459 & 0.0995 & 0.1502 & 0.0555 & 0.0900\\ \hline
 \multicolumn{9}{c}{\yelp}
  \\ \hline
 \pop 		  & 0.0023 & 0.0022 & 0.0018 & 0.0008 & 0.0039 & 0.0092 & 0.0017 & 0.0051\\ 
 \bbpr		  & 0.0097 & 0.0082 & 0.0067 & 0.0033 & 0.0139 & 0.0342 & 0.0062 & 0.0188\\
 \warp		  & 0.0139 & 0.0112 & 0.0088 & 0.0047 & 0.0184 & 0.0437 & 0.0084 & 0.0247\\ 
 CBOW  		  & 0.0165 & 0.0125 & 0.0097 & 0.0059 & 0.0219 & 0.0499 & 0.0100 & 0.0283\\
RNN                & \textit{0.0183} & \textit{0.0138} & \textit{0.0104} & \textit{0.0069} & \textit{0.0248} & \textit{0.0551} & \textit{0.0113} & \textit{0.0313}\\
 \awarp		  & 0.0142 & 0.0117 & 0.0098 & 0.0046 & 0.0193 & 0.0430 & 0.0087 & 0.0249\\	
 HA-RNN          & \textbf{0.0205} & \textbf{0.0154} & \textbf{0.0117} & \textbf{0.0080} & \textbf{0.0281} & \textbf{0.0638} & \textbf{0.0131} & \textbf{0.0357}\\ 
HA-RNN*  & 0.0221 & 0.0164 & 0.0123 & 0.0083 & 0.0301 & 0.0665 & 0.0140 & 0.0377 \\ \hline
\end{tabular}
\end{table*}
We compare HA-RNN with other state-of-the-art models and report the results in Table \ref{t:stoa}. We interpret the table as follows. First, attribute embedding is vital in \xing---the scores of \awarp~and~\lstm~are significantly better than those of~\warp,~\cbow, and RNN. Recommendation does benefit from sequence approaches as RNN performs better than \warp; \lstm~does better than \awarp. Second, sequence approach is critical in \yelp~as we see \cbow, RNN, and HA-RNN clearly beat \warp~and~\awarp. Finally, the best accuracy is achieved when HA-RNN combines both attribute embedding and sequence modeling. The improvements are significant. On~\xing~dataset, CA-RNN \textit{single model} relatively improves Precision@2, Recall@30, MAP, and NDCG by 29\%, 25\%, 36\%, and 30\%; on~\yelp~dataset, the improvements are 12\%, 16\%, 16\%, and 14\%.

\begin{table*}
\centering
\caption{Recommendation accuracy comparisons of models with different sequence modeling capabilities.}
\label{t:seq}
\begin{tabular}{ccccccccc} \hline
Methods & P@2 & P@10 & P@30 & R@2 & R@10  & R@30 & MAP & NDCG \\ \hline
\multicolumn{9}{c}{\xing}
  \\ \hline
HMF     		& 0.0362 & 0.0194 & 0.0109 & 0.0325 & 0.0749 & 0.1163 & 0.0398 & 0.0674\\
 NHMF 		& 0.0331 & 0.0190 & 0.0112 & 0.0272 & 0.0734 & 0.1227 & 0.0359 & 0.0653 \\ 
HA-SG		& 0.0328 & 0.0191 & 0.0113 & 0.0264 & 0.0727 & 0.1026 & 0.0355 & 0.0651 \\
HA-CBOW	& 0.0377 & 0.0208 & 0.0119 & 0.0331 & 0.0842 & 0.1355 & 0.0425 & 0.0741 \\
HA-RNN	& 0.0522 & 0.0241 & 0.0128 & 0.0451 & 0.0962 & 0.1453 & 0.0540 & 0.0873\\  \hline
\multicolumn{9}{c}{\yelp}
  \\ \hline
HMF		& 0.0142 & 0.0117 & 0.0098 & 0.0046 & 0.0193 & 0.0430 & 0.0087 & 0.0249\\
NHMF  		& 0.0148 & 0.0117 & 0.0091 & 0.0051 & 0.0206 & 0.0471 & 0.0093 & 0.0265\\ 
HA-SG		& 0.0151 & 0.0122 & 0.0094 & 0.0047 & 0.0205 & 0.0477 & 0.0090 & 0.0266 \\
HA-CBOW	& 0.0174 & 0.0136 & 0.0104 & 0.0061 & 0.0236 & 0.0540 & 0.0106 & 0.0304 \\
HA-RNN	& 0.0205 & 0.0154& 0.0117 & 0.0080 & 0.0281 & 0.0638 & 0.0131 & 0.0357 \\  \hline
\end{tabular}
\end{table*}

\subsection{Other Sequential Recommendation Models}

We want to empirically study the effect of sequence modeling on recommendation performance. In order to do that, the same embedding techniques in HA-RNN are applied to skip-gram and CBOW models (we call the new models HA-SG and HA-CBOW). In this way, we assume the new models have similar capabilities in incorporating attributes but differ in their abilities in capturing sequential dependencies. The results are shown in Table \ref{t:seq}. We make three observations below.

First, non-sequence models HMF and NHMF have similar performances. Second, HA-CBOW clearly beats HMF and NHMF on both datasets although it has a similar bi-linear model formulation as in HMF and NHMF. It shows HA-CBOW does benefit from its sequence-based training. Finally, HA-RNN outperforms HA-CBOW significantly. We attribute it to its stronger ability in sequence modeling and its flexibility.

\subsection{Learned Attribute Embedding}

\begin{figure}[t]
\caption{2-D visualization of learned embedding of \xing's categorical attributes by t-SNE. Different colors represent different attribute types. Some local regions are zoomed in. Best viewed in color.}\label{pic:emb}
\includegraphics[width=0.99\columnwidth]{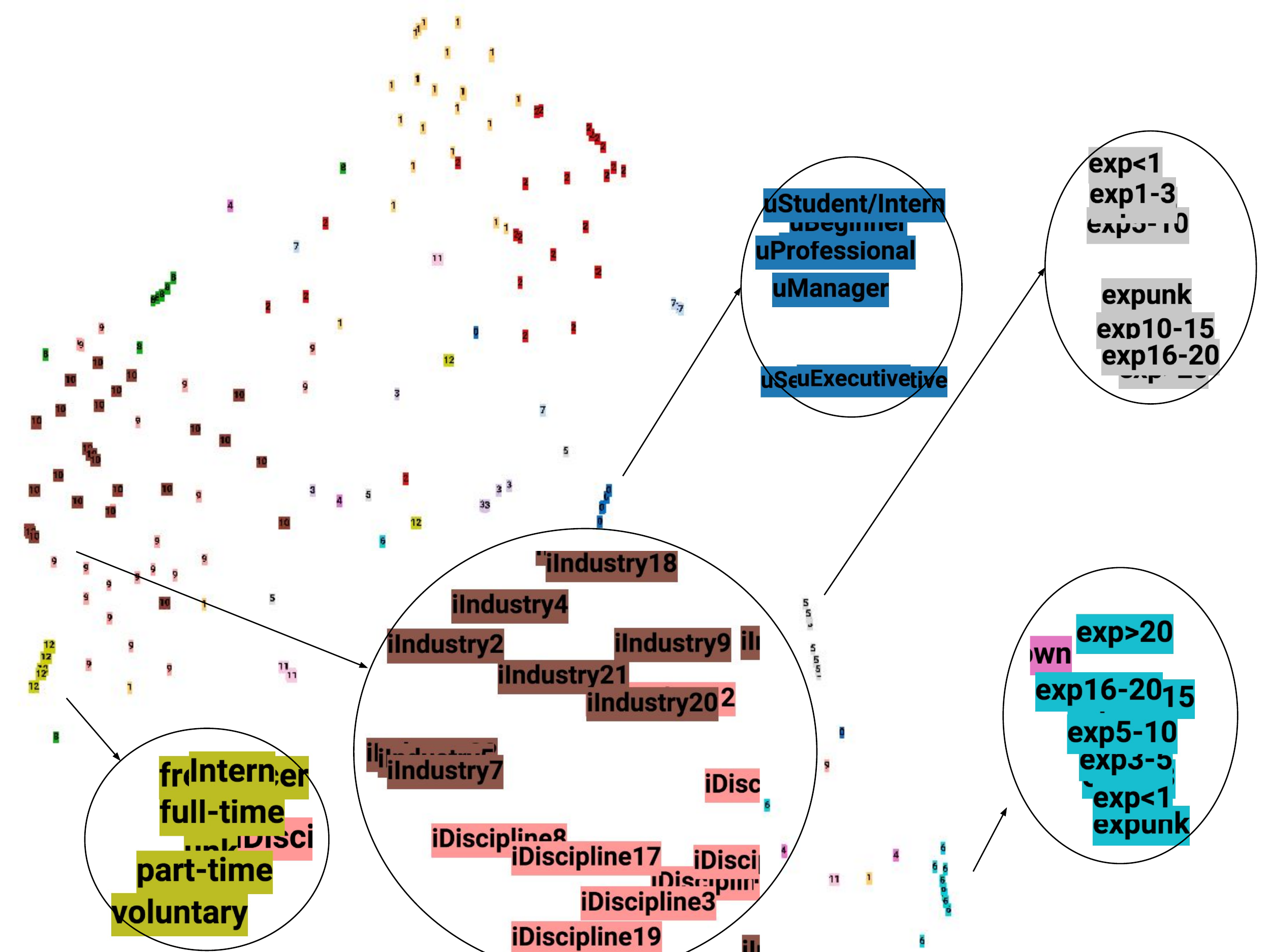}
\end{figure}

We give our qualitative analysis on the embedding learned by~\lstm. We take the learned attribute embedding and plot the 2-D visualization (Fig.~\ref{pic:emb}). On~\xing, we observed an interesting hierarchical clustering effect of categorical attribute embedding. First, the same type of attribute embedding tends to cluster. For example, user career levels such as \textit{student/intern}, \textit{manager}, \textit{executive}, etc., form a cluster; \textit{executives}, and \textit{senior executive} understandably stand a bit farther from the rest. Second, embedding clusters across types tends to stay close when they have close semantic meaning. For example, (anonymous) attributes of industry and discipline types are closely related and thus have close distances. This is encouraging since we started from heterogeneous attributes and managed to infer their semantic structures.

On \yelp, we take a look at the learned embedding of multi-hot attribute ``categories.''\footnote{In \xing, we do no analyze the multi-hot attribute embedding because the tokens are anonymized there.} We see that embedding reflects the correlation of business types in the reviews. For example, we locate the top nearest neighbors of \textit{juice bars \& smoothies} (distances are computed in the embedding space): \textit{macarons}, \textit{cardio classes}, \textit{meditation centers}, \textit{cafes}, \textit{wine tasting room} .... This suggests that people who review a business of type \textit{juice bars \& smoothies} should likely review businesses of type \textit{cardio classes} or \textit{wine tasting room}. The information might be used in cases where a new business is  recommended to a Yelp customer. Some other examples (including those from \xing) are listed in Table~\ref{t:emb_dis}.

\begin{table}
\centering
\caption{Examples of attributes and their nearest neighbors in the embedding space (from \yelp~and \xing).}
\begin{tabular}{|c|c|} \hline
Attribute& Nearest Neighbors \\ \hline \hline
korean & vietnamese, taiwanese, japanese.... \\ \hline 
children's  & baby gear \& furniture, children's museums, \\
clothing & maternity wear, cosmetics \& beauty supply.... \\ \hline \hline
bachelor & master, phd, student/intern, industry6.... \\ \hline
exp$<$1 & exp1-3, exp3-5, exp5-10, exp10-15.... \\ \hline
\end{tabular}
\label{t:emb_dis}
\end{table}

\begin{figure*}[t]
\caption{Results on~\xing~when increasing proportion of data is used (proportion \{0.2, 0.44, 0.76, 1.0\}). Sequence model \lstm~(red) performance improves steadily while that of non-sequence model \nhmf~(blue) does not.}\label{pic:scale}
\centering
\subfigure[Precision@5]
{\label{pic:p5}\includegraphics[width=0.48\columnwidth]{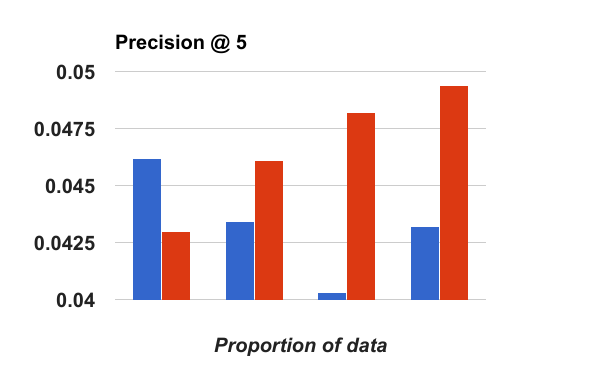}
}
\hspace{-0.4cm}
\subfigure[Recall@30]
{\label{pic:r5}\includegraphics[width=0.48\columnwidth]{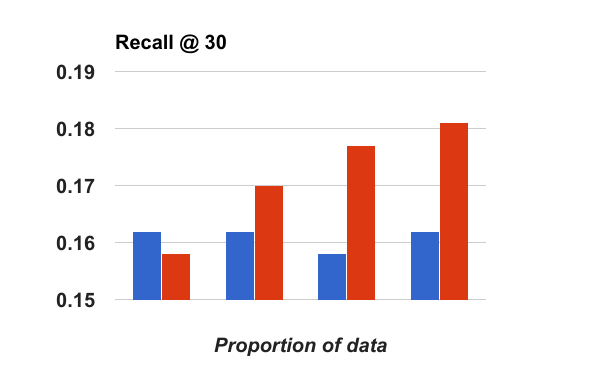}
}
\hspace{-0.4cm}
\subfigure[NDCG]{\label{pic:NDCG}
\includegraphics[width=0.48\columnwidth]{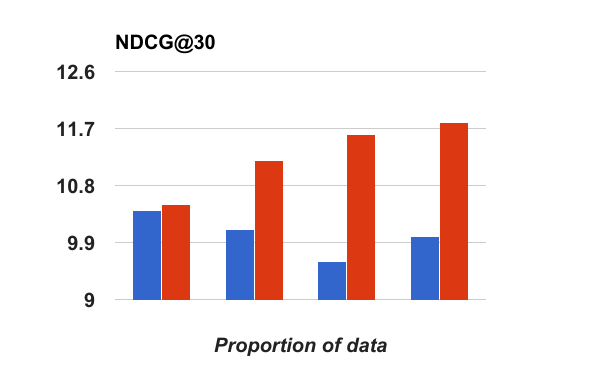}
}
\hspace{-0.4cm}
\subfigure[SCORE]{\label{pic:score}
\includegraphics[width=0.48\columnwidth]{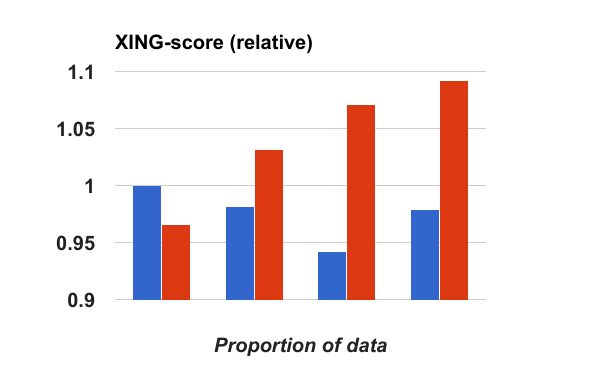}
}
\end{figure*}

\subsection{Sequence Modeling Studies}
The observation in Table \ref{t:stoa} that \lstm~outperforms \awarp~and RNN suggests the benefits in the explicit modeling of sequential properties of both item and attributes. We are further interested in the following two questions.

\textit{Should we sample items from item sequences or not during model training?} When applying sequence modeling to the recommendation problem, we implicitly assume that sequence or order provides additional information beyond that provided by item frequency alone. Is this a valid assumption and how strong is the sequential dependency? In the modeling perspective, \lstm~has two major differences compared to \awarp: explicit sequence modeling and flexible non-linear function mapping. It is possible that the advantages of \lstm~mainly come from the latter. Meanwhile, data augmentation by sampling is widely used as a regularization. It may help improve the performance by sampling items.

\textit{What if dataset size changes?} The current experiments are performed on two fixed-size datasets. We are wondering if the better performance of sequence models is an artifact of small dataset sizes? In other words, does the non-sequence model catch up with sequence models when more observations are given?

To investigate these two questions, and in general how item sequences influence the performance of  \lstm, we conducted additional experiments on \xing~and report two of our findings below.\footnote{Scores on the development set are used in these experiments. We find they are strongly correlated with test scores.}

\textit{Sequence vs. frequency}
In the new experiments with \lstm, we generated new training data through sampling subsequences in which items in a user's item sequence were dropped out with certain probability. The sampling procedure is repeated to augment item sequence. For example, an item sequence $\{ i_1, i_2, i_3, i_4, i_5\}$ might now become three subsampled sequences: $\{i_1, i_3, i_4\}$, $\{i_2, i_3, i_5\}$, and $\{i_1, i_2, i_4, i_5\}$. New sequences were used in the training, but on average item appearance frequency tended to be unchanged in data with more and more augmented subsequences due to unbiased sampling. 

Experimental results with the new generated training data are shown in Table \ref{t:sample}. ``Original'' denotes dataset based on the complete sequence (i.e., no sampling augmentation). $x_N$ denotes the manipulated data set obtained by sampling and augmenting subsequence proportional to $N$ times. First, increasing subsequence sampling leads to decreasing scores (from $x_1$ to $x_8$); second, the original data set (full sequences) gives the best score. These results suggest that it is harmful to break item sequence even though the item frequency is maintained the same. This verifies our assumption that item sequences indeed provide additional information than frequency alone.

\begin{table}
\centering
\caption{\lstm~relative scores on original and ``manipulated'' \xing~datasets. ``Original'' denotes the complete sequence. $x_N$ denotes the manipulated data set obtained by randomly sampling and augmenting subsequence proportional to $N$ times.}
\label{t:sample}
\begin{tabular}{|c|c|c|c|c|c|}  \hline
Sampling & Original & $x_1$ & $x_2$ & $x_4$ & $x_8$ \\ \hline
Score & 1.0 & 0.97 & 0.94 & 0.86 & 0.84 \\ \hline
\end{tabular}
\end{table}

\textit{Performance vs. sequence number}
Another way to test our assumption is to increase (or decrease) the number of observed sequences and to compare performance between non-sequence and sequence models. While both non--sequence and sequence models observe the same total number of user-item pairs, sequence models might have a chance to extract more useful information from the data if our assumption is true.

We train models \nhmf~and \lstm~on \xing~and still evaluate on \textit{target users}. We gradually increase the training observations from those of \textit{target users} to those of a super-set of users until all users are included. The percentages of total interaction used for four experiments are 20\%, 44\%, 76\%, and 100\%, respectively.

Results are reported in Figure \ref{pic:scale}. We see that across scores returned by different metrics,\footnote{`SCORE' is as used in RecSys Challenge 2016 \cite{abel2016recsys}.} \lstm~improves steadily when the data scale increases (more sequences observed). On the contrary, \nhmf~in general does not improve. This can be explained by our assumption that useful information is encoded in the sequence: recommendation accuracy improves when more sequences are observed. This is not the case when more independent user-item pairs are observed. Our proposed  \lstm~approach successfully utilizes this helpful information and benefits from that. From another perspective, it suggests that sequence models may even be better suited when we have larger-scale recommendation from implicit feedback. Based on this, we would advocate the use of sequence modeling in large scale recommendation setting.

%% file: conclude.tex
\section{Conclusion}

In this paper we explore the effectiveness of combining heterogeneous attribute embedding and sequence modeling in recommendation with implicit feedback. To this end, we build a neural network framework to incorporate attributes and apply sequence models to embed attributes and to perform recommendation. Through empirical studies on four large-scale datasets, we find that rich discriminative information is encoded in heterogeneous attributes and item sequences. By combining attribute embedding and flexible sequence models, we are able to capture  information and improve recommendation performance.